\documentclass{PoS}
\usepackage{booktabs,multirow,tabularx}
\usepackage{fmtcount}

\newcommand\CutTools{{\sc\small CutTools}}
\newcommand\MadLoop{{\sc\small MadLoop}}
\newcommand\ML{{\sc\small ML}}
\newcommand\MadGraph{{\sc\small MadGraph}}
\newcommand\FeynRules{{\sc\small FeynRules}}
\newcommand\MGF{{\sc\small MG5}}
\newcommand\aMCatNLO{{\sc\small aMC@NLO}}
\newcommand\HELAS{{\sc\small HELAS}}
\newcommand\qpart{q^\star}
\newcommand\qbpart{\bar{q}^\star}
\newcommand\sss{\scriptscriptstyle}
\newcommand\as{\alpha_{\sss S}}

\title{New developments in {\sc\huge MadLoop}}

\ShortTitle{Public version of {\sc\small\emph MadLoop }}
\author{\speaker{Valentin Hirschi}\\
        Institute for Theoretical Physics, EPFL, Lausanne, Switzerland\\
        E-mail: \email{valentin.hirschi@epfl.ch}}

\abstract{In this talk, the recent developments of \MadLoop\ are presented. \MadLoop\ automates the computation of one-loop QCD corrections to an arbitrary scattering process in the Standard Model. I first review the current version of the code which has been made \emph{public}  through the \aMCatNLO\ webpage. In the second part, progress in the implementation of \MadLoop\ within the \MGF\ framework is presented along with the preliminary speed benchmarks for a few selected massless QCD processes.}

\FullConference{ 10th International Symposium on Radiative Corrections (Applications of Quantum Field Theory to Phenomenology) - Radcor2011\\
September 26-30, 2011\\
Mamallapuram, India}

\begin{document}

\section{{\sc\large MadLoop}\ in a nutshell}

Recently, several programs~\cite{Cullen:2011ac,Bevilacqua:2011xh,Campbell:2010ff, Hirschi:2011pa, Berger:2008sj, Berger:2008ag} for the computation of one-loop corrections in QCD have been developed thanks to a new generation of tools based on numerical~\cite{Ossola:2007ax, Heinrich:2010ax} or semi-analytical~\cite{Cullen:2011ac,Bevilacqua:2011xh,Campbell:2010ff, Berger:2008sj} reduction mechanisms. \MadLoop~\cite{Hirschi:2011pa} exploits the OPP~\cite{Ossola:2008zza} method as implemented in \CutTools~\cite{Ossola:2007ax} and aims at  the full automation of arbitrary 1-loop computations. In its original version \MadLoop\ is an independent {\tt{C++}} code interfaced to \MadGraph 4~\cite{stelzer-1994-81} for tree-diagram generation that writes out a fortran code computing the amplitude as requested by a user. Recently, {\tt python} version five of \MadGraph, dubbed \MGF~\cite{Alwall:2011uj}, has been released and the possibility of rewrting  \MadLoop\  in this framework has arisen. We refer to this new version as \MadLoop5, as opposed to the original \MadLoop4 which is now part of the \aMCatNLO\ framework and has already sucessfully provided virtual amplitudes for several calculations and LHC phenomenological applications~\cite{Frederix:2011zi, Frederix:2011qg,Frederix:2011ss, Frederix:2011ig}.

What follows gives a quick description of the \MadLoop\ program.

\subsection{Program overview}

The user process $\{n_i\}\rightarrow\{n_f\}$ is specified in an order file compliant with the \emph{LesHouches} standards~\cite{Boos:2001cv}, as well as the model name and the desired QED order.  The Born diagrams are generated along with the necessary counterterms (UV renormalization and the so-called R2 terms~\cite{Draggiotis:2009yb}). The loop topologies are constructed starting from their corresponding L-cut diagrams, which are obtained by cutting\footnote{These cuts have nothing to do with unitarity cuts and this is why they are referred to as \emph{L-cuts}, standing for Loop cuts.} the loop at a given internal line. L-cut diagrams are obtained by generating the tree-level diagrams of the process $\{n_i\}\rightarrow\{n_f\}+\{\qpart,\,\qbpart\}$, where the $\qpart$ are the L-cut particles, namely those that can run in the loop. For QCD corrections, it is enough to consider for $\qpart$ the gluon and all the quarks. 

As loops can be cut at as many places as there are loop propagators times two (they can be read along two opposite directions), many L-cut diagrams correspond to the same loop diagram once the two L-cut particles are sewed back together. The first task of \MadLoop\ is to efficiently identify double counting by associating to each L-cut diagram a tag which is unique to each loop diagram contributing to the process.

Accounting for the color structure of the amplitude is a straight-forward extension of what is done at the tree-level except for the fact that in this case virtual-Born interference terms must be considered. The computation of the L-cut diagrams is carried on exactly as for tree-level processes through chains of \HELAS~\cite{Helas:} calls. However, in order to provide the integrand necessary for \CutTools\ to numerically evaluate the loop contribution, some modifications to the tree-level output are necessary:
\begin{itemize}
\item
The momentum flowing in the loop is complex.
\item
Only the numerator of the propagator of the loop particles must be computed.
\item
The Lorentz trace of the loop is realized by summing over the arbitrary polarization vectors of the L-cut particles, for which an orthogonal basis is used.
\end{itemize}
To assess the sanity of the result, two checks are automatically performed: the coefficient of the double pole of the virtual amplitude is compared against its universal analytical expression taken from eq. B.2 of ref.~\cite{Frederix:2009yq} and, in case a gluon or a photon are present in the external state, gauge invariance is tested by calculating the 1-loop amplitude with a longitudinal polarization.

A key issue in automatic computations is the detection and efficient management of possible numerical instabilities.
Phase space points for which \CutTools\ detects that some loops are numerically unstable, called UPS points further then, do undertake a "rescue" process at the helicity level. If the kinematical configuration of the UPS is slightly modified, it is likely that it will fall back in a stable regime. The following \emph{shift} is applied\footnote{The shift of Eq.~\ref{shift} is performed along the z-axis but other directions are also considered.}:
\begin{equation}
k_i^3\;\longrightarrow\;(1+\lambda_\pm) k_i^3\,.
\label{shift}
\end{equation}
The transverse momenta are left invariant, and the energy components are adjusted so to impose on-shellness and four-momentum conservation. If, for at least two of these modified phase-space points, the result for the originally unstable loop is flagged stable, the evaluation of the finite part of the virtual returned by \MadLoop\ is based on the average of the stable evaluations of the two shifted phase-space points, normalized by their respective Born amplitude. The uncertainty due this procedure is deduced from the distance between these two shifted results. Among all the processes presented in ref.~\cite{Hirschi:2011pa} only a couple have a fraction of UPS larger than $10^{-4}$ of which 99.9\% are recovered to stability after shifts of magnitude $|\lambda|=0.01$ or $|\lambda|=0.05$ at most. For the very few UPS still unstable after the shifts, \MadLoop\ returns a value based on the median of the distribution of all previous stable evaluations.
This rescue system can guarantee an uncertainty due to UPS much smaller than the Monte-Carlo one for all processes presented in~\cite{Hirschi:2011pa}, where the interested reader can also find more details on the rescuing method.

\section{Public use of {\sc\large MadLoop4}}
\label{publicMadLoop}

The use of {\MadLoop}4 has recently been made public\footnote{The user still needs to register on the webpage before using this tool. The registration process, automatic and fast, is only here to keep track of the different users of the tool.} through the website \emph{amcatnlo.web.cern.ch} under the section 'Compare with MadLoop'. For a wide range of processes in the Standard Model, the user can obtain the numerical evaluation of the finite part and poles of the virtual amplitude squared against the Born one and summed over color and helicity. The only inputs required by the user are:
\begin{itemize}
\item
The process, specified using the \MadGraph\ naming conventions and compliant with the \MadLoop 4 limitations (see further).
\item
The QED coupling order of the process.
\item
Wether to consider the bottom quark massive or massless.
\end{itemize}
At this stage, the input process must still respect all \MadLoop 4 constraints listed below:
\begin{itemize}
\item
The process must have a contribution at tree-level (Born level).
\item
The Born Feynman diagrams contributing must not comprehend any 4-gluon vertex.
\item
All Born contributions must factorize the same powers of $\as$ and $\alpha$.
\item
At least one quark must be present in the external states of the input process.
\end{itemize}
The conventions adopted by \MadLoop 4 for the computation are:
\begin{itemize}
\item
UV renormalization is included, in the $\overline{MS}$ scheme for massless modes and using zero-momentum subtraction for the massive ones. See appendix B.2 of ref.~\cite{Hirschi:2011pa} for the definition of the UV counterterms employed.
\item
The widths of all particles are set to zero as \MadLoop 4 does not work in the complex mass scheme.
\item
The identical final state particle factor is included.
\end{itemize}
The result $V$ is given in the CDR regularization scheme and in the form of three coefficients\footnote{Notice that if any contributing Born diagram contains a massive virtual quark (i.e. on an internal line of the Feynman Diagram), only the coefficient $c_0$ and $c_{-2}$ are correctly computed and $c_{-1}$ will not be provided.} $c_0$, $c_{-1}$ and $c_{-2}$ defined as follows:
\begin{equation}
V=\frac{(4\pi)^\epsilon}{\Gamma(1-\epsilon)}
\left(\frac{c_{-2}}{\epsilon^2}+\frac{c_{-1}}{\epsilon}+c_0\right)\,.
\label{Vexpr}
\end{equation}
The model parameters used can be found on the \aMCatNLO\ \MadLoop\ check webpage.


\section{Recent progresses in {\sc\large MadLoop5}}

As advertised in the introduction, \MadLoop\ has recently been implemented in the \MGF\ framework. Thanks to the great flexibility of \MGF, \MadLoop 5 already brings new improvements.

\begin{figure}[ht]
\begin{center}
\begin{minipage}{1.0\linewidth}
\centering
\includegraphics[width=0.8\linewidth]{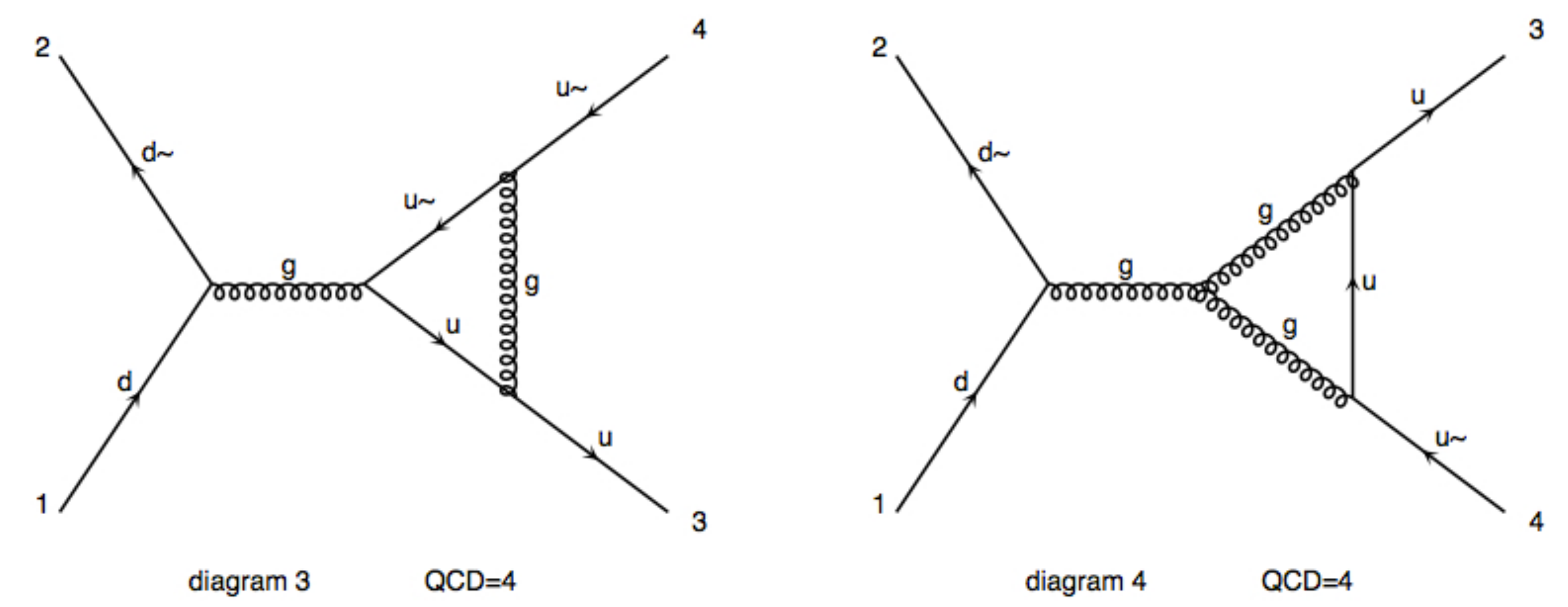}
\caption{Two examples of loop Feynman diagrams drawn by \MadLoop 5 for the process $\;d\;\bar{d}\rightarrow\;u\;\bar{u}$.}
\label{loop_diagrams}
\end{minipage}\hfill
\end{center}
\end{figure}

Firstly, \MadLoop 5 generates loop-diagrams much faster thanks to the more efficient \MGF\ algorithm for tree-diagram generation. Moreover, all \MadLoop 4 limitations listed in section~\ref{publicMadLoop} are relieved. The cumbersome computation of the four gluons R2 counterterm is made possible by exploiting the UFO~\cite{Degrande:2011ua} model format which allows for vertex definitions with entangled color and Lorentz structures.
As a secondary upgrade, the loop diagrams themselves can now be drawn and not only the L-cut diagrams as it was the case in \MadLoop 4.  Figure~\ref{loop_diagrams} shows a snapshot of the drawings of loop diagrams contributing to the $\;d\;\bar{d}\rightarrow\;u\;\bar{u}$ process. Finally, a more efficient caching of the HELAS wavefunctions computed during the evaluation of a phase-space point improves running time.

In the near future, \MadLoop 5  will also feature other new capabilities. This includes the ability of working in the complex mass scheme and to have massive electroweak bosons running in the loop. For this to be useful, mixed order perturbation corrections will be supported\footnote{\MGF\ can already generate the relevant loop for processes for which the user includes the perturbation of different coupling orders, like QCD and QED.}. 
To more efficiently handle numerically unstable phase-space points, \MGF\ will be able to provide the loop integrand to \CutTools\ in multiple-precision when needed. Finally, decay chains will be specified and generated in the exact same way as they are now for tree-level processes in \MGF.
In a further future, \MadLoop 5 will be capable of working in any BSM models obtained directly from \FeynRules~\cite{Christensen:2008py}.
\\
First results from \MadLoop 5 are already available for the massless QCD model with 2 light flavor quarks. Table~\ref{tab:benchmark} shows a comparison between results obtained with \MadLoop 4 and \MadLoop 5 for chosen processes. The five- and six-gluon amplitude were successfully cross-checked against the {\sc\small NGluon}~\cite{Badger:2010nx} code. Drastic improvement in generation time and output file size is obtained while running time is only moderately improved.

\addtocounter{footnote}{1}
\begin{table}
\begin{center}
\begin{tabular}{|l|c|c|c|c|c|c|}
\hline
\multirow{2}{*}{Process} & \multicolumn{2}{c|}{Generation time} & \multicolumn{2}{c|}{Output size} & \multicolumn{2}{c|}{Running time} \\
& \ML5 & \ML4 & \ML5 & \ML4 & \ML5 & \ML4 \\
\hline
$d\bar{d}\rightarrow u \bar{u}$& 8.8s & 5.4s  & 208 Kb & 268 Kb & 0.0088s & 0.0094s \\
$d\bar{d}\rightarrow d \bar{d} g$& 17s & 104s  &   424 Kb & 1.7 Mb  &  0.64s & 0.74s  \\
$d\bar{d}\rightarrow d \bar{d} u \bar{u}$ & 23s & 2094s  & 552 Kb & 3.3 Mb  &  1.93 s & 2.34s  \\
$g g \rightarrow g g g$&   71 s & X  &  1.5 Mb & X  & 30s & X  \\
$g g \rightarrow g g g g$&   37m60s$^{\decimal{footnote}}$ & X  &  27 Mb & X  & 1h12m & X \\
\hline
\end{tabular}
\end{center}
\caption{\label{tab:benchmark}
Speed benchmark comparing results for chosen processes obtained with \MadLoop 5 and \MadLoop 4 on a machine with a dual 2.8 Ghz Intel Core CPU. The running time refers to the computation of the loop amplitude squared (against the Born) {\it summed over color and helicity configurations} for a single phase-space point. In particular, only the vanishing helicity configurations are filtered out, not the vanishing loops for specific helicity configurations. \CutTools\ is allowed to re-perform the reduction in multiple-precision when needed. The reference model used here is massless QCD with two light-flavored quarks. The output file size includes the HELAS source only for \MadLoop 5 as  it is process independent for \MadLoop 4 (no optimization is performed to get rid of the unused subroutines). The five- and six-gluon amplitudes are marked with a cross for \MadLoop 4 as they are not available.
}
\end{table}
\footnotetext[\value{footnote}]{Compilation time dominates here as gfortran takes 3h30 to compile the output code.}

\newpage
\section{Conclusion}
In the first section, the public version of \MadLoop4 with the details of its computational setup and limitations were presented. Applications 
to physics studies and its use as robust benchmark for other private computations have already proven to be useful.

In the last section, the preliminary results for the \MadLoop\ implementation in \MGF\ were shown along with the new features already implemented and yet to come. This paves the way to a whole set of new applications using \MGF\ for LHC phenomenology.

\section{Acknowledgements}

I am indebted to Marco Zaro for his great help in setting up the webpage for public use of \MadLoop4 and to Olivier Mattelaer for the implementation of the Feynman loop diagram drawing in \MadLoop 5. I would also like to thank Johan Alwall for many useful discussions and his insightful ideas on the \MadLoop 5 structure. Finally, I am grateful to Fabio Maltoni for the thorough proofreading of this proceeding and to all members of the \aMCatNLO\ group in general without whom this work would not have been possible.

\bibliographystyle{JHEP}
\bibliography{Library}


\end{document}